\documentclass[conference]{IEEEtran}

\ifCLASSINFOpdf
   \usepackage[pdftex]{graphicx}
   \DeclareGraphicsExtensions{.pdf,.jpeg,.png}
\else
\fi

\usepackage{tabularx,calc}

\usepackage{tikz,amsmath, amssymb,bm,color}
\usepackage{pgfplots}
\usepackage{environ}
\usepackage{ifthen}
\makeatletter
\newsavebox{\measure@tikzpicture}
\NewEnviron{scaletikzpicturetowidth}[1]{%
	\def\tikz@width{#1}%
	\begin{lrbox}{\measure@tikzpicture}%
		\BODY
	\end{lrbox}%
	\pgfmathparse{#1/\wd\measure@tikzpicture}%
	\BODY
}
\makeatother
\newboolean{compileforpublish}
\setboolean{compileforpublish}{true}
\newboolean{isaccepted}
\setboolean{isaccepted}{true}
\ifthenelse{\boolean{isaccepted}}
{%if
	\newcommand\copyrighttext{%
		\footnotesize \parbox[t]{.11\textwidth}{\copyright{} 2016~IEEE.} \parbox[t]{.89\textwidth}{Personal use of this material is permitted. Permission from IEEE must be obtained for all other uses, in any current or future media, including reprinting/republishing this material for advertising or promotional purposes, creating new collective works, for resale or redistribution to servers or lists, or reuse of any copyrighted component of this work in other works.}}
}
{%else
	\newcommand\copyrighttext{%
		\footnotesize \centering This work has been submitted to the IEEE for possible publication.\\ Copyright may be transferred without notice, after which this version may no longer be accessible.}
}

\newcommand\copyrightnotice{%
	\ifthenelse{\boolean{compileforpublish}}
	{
		\begin{tikzpicture}[remember picture,overlay]
		\node[anchor=south,yshift=10.5pt] at (current page.south) {\parbox{\dimexpr\textwidth-\fboxsep-\fboxrule\relax}{\copyrighttext}};
		\end{tikzpicture}%
	}
}

\renewcommand{\IEEEtitletopspaceextra}{20pt}

\begin{document}
\title{Identification of Potential Hazardous Events for an Unmanned Protective Vehicle}

\author{\IEEEauthorblockN{Gerrit Bagschik, Andreas Reschka, Torben Stolte and Markus Maurer}
\IEEEauthorblockA{Institute of Control Engineering\\
Technische Universit\"at Braunschweig\\
Braunschweig, Germany\\
Email: \{bagschik, reschka, stolte, maurer\}@ifr.ing.tu-bs.de}
}
\maketitle%
\copyrightnotice%

\begin{abstract}
The project \emph{Automated Unmanned Protective Vehicle for Highway Hard Shoulder Road Works} (aFAS) aims to develop an unmanned protective vehicle to reduce the risk of injuries due to crashes for road workers.
To ensure functional safety during operation in public traffic the system shall be developed following the ISO~26262~standard.
After defining the functional range in the item~definition, a hazard analysis and risk assessment has to be done.
The ISO~26262~standard gives hints how to process this step and demands a systematic way to identify system hazards.
Best practice standards provide systematic ways for hazard identification, but lack applicability for automated vehicles due to the high variety and number of different driving situations even with a reduced functional range.
This contribution proposes a new method to identify hazardous events for a system with a given functional description.
The method utilizes a skill graph as a functional model of the system and an overall definition of a scene for automated vehicles to identify potential hazardous events.
An adapted Hazard and Operability Analysis approach is used to identify system malfunctions. 
A combination of all methods results in operating scenes with potential hazardous events.
These can be assessed afterwards towards their criticality.
A use case example is taken from the current development phase of the project aFAS.

\end{abstract}

\section{Scope of work}

\begin{figure}[h]
	\begin{center}
	\includegraphics[width=5cm]{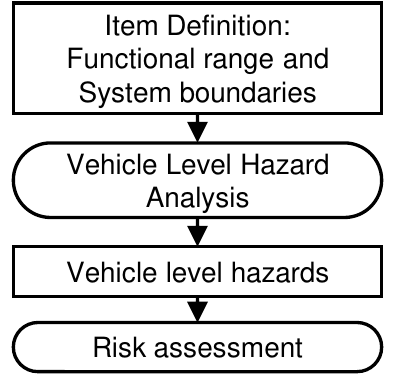}    % The printed column width is 8.4 cm.
	\caption{Concept phase of ISO~26262~standard (simplified) with process steps (rounded) and work products (cornered) \cite[Part 3]{ISO_26262_2001}} 
	\label{fig:conceptphase}
	\end{center}
\end{figure}

The project \emph{Automated Unmanned Protective Vehicle for Highway Hard Shoulder Road Works} (aFAS\footnote{German abbreviation for \emph{Automatisch fahrerlos fahrendes Absicherungsfahrzeug f{\"u}r Arbeitsstellen auf Autobahnen}}) aims to develop an unmanned protective vehicle to reduce the risk of injuries by crashes for road workers.
The unmanned protective vehicle follows a leading vehicle in a defined distance on the hard shoulder of a highway without a safety driver or human supervision.
On- and off-ramps are passed in very close distance to the leading vehicle.
A detailed outline of the project aFAS and the main objectives are described in \cite{stolte_towards_2015}.
Despite the operation on a hard shoulder of a highway this project aims to show the first operation of an unmanned vehicle in public traffic on German roads.
Due to safety criticality the system shall be developed applying the ISO~26262~standard \cite{ISO_26262_2001} for ensuring functional safety.
The project is currently in the concept phase of the reference development process proposed by the ISO~26262~standard, which is shown in Figure~\ref{fig:conceptphase}.
After defining the range of features and the functional system boundaries in the \emph{item~definition}, the system is inspected with regards to criticality during operation and the risk to other traffic participants in the \emph{hazard analysis and risk assessment} (HARA).
The first step of the HARA is the scene analysis (for a definition of the term scene see \cite{ulbrich_scene_2015} and Section \ref{subsec:scenes}), which shall identify operational scenes where malfunctioning behavior of the system can lead to mishaps which are called potential hazardous events.
The ISO~26262~standard uses the term situation which equals the term scene defined in \cite{ulbrich_scene_2015} and is called \emph{scene} for this contribution.
%These scenes are called potential hazardous events.
These scenes thus describe the correct use and the misuse of the item in a foreseeable way.
A methodology for the identification of such scenes is the main focus of this contribution. \\
According to the ISO~26262~standard, the hazardous events ``shall be determined systematically by using adequate techniques'' and ``based on the item's functional behaviour; therefore, the detailed design of the item does not necessarily need to be known.'' \cite[Part~3]{ISO_26262_2001}
The proposed techniques are brainstorming, checklists, quality history, Failure Mode and Effects Analysis (FMEA) and field studies \cite[Part~3]{ISO_26262_2001}.
Due to the variety of operational situations, non-structured brainstorming and checklists do not seem to be a legitimate way for covering all relevant situations.
%\textcolor{red}{Rechenbeispiel}
A quality history or field studies are not available because the \emph{unmanned protective vehicle} will be the first of its type, therefore examples of current protective vehicles can be used as a base.
It has to be noticed that the major difference that no human operator is available in the vehicle leads to other (more or different) hazardous events.
The classical FMEA needs a detailed design of the analyzed system, which per definition is not available in the concept phase.
A review of the ISO~26262~standard by Van Eikema Hommes points out that ``the lack of guidance on hazard identification and elimination hinders the standards ability to sufficiently provide safety assurance'' \cite{hommes_review_2012}.
For these reasons this contribution proposes a novel systematic method for the identification of  potential hazardous events for the unmanned protective vehicle.

\section{Understanding of the Term Hazard for Automated Vehicles}

\begin{figure}[h]
	\begin{center}
	\includegraphics[width=0.48\textwidth]{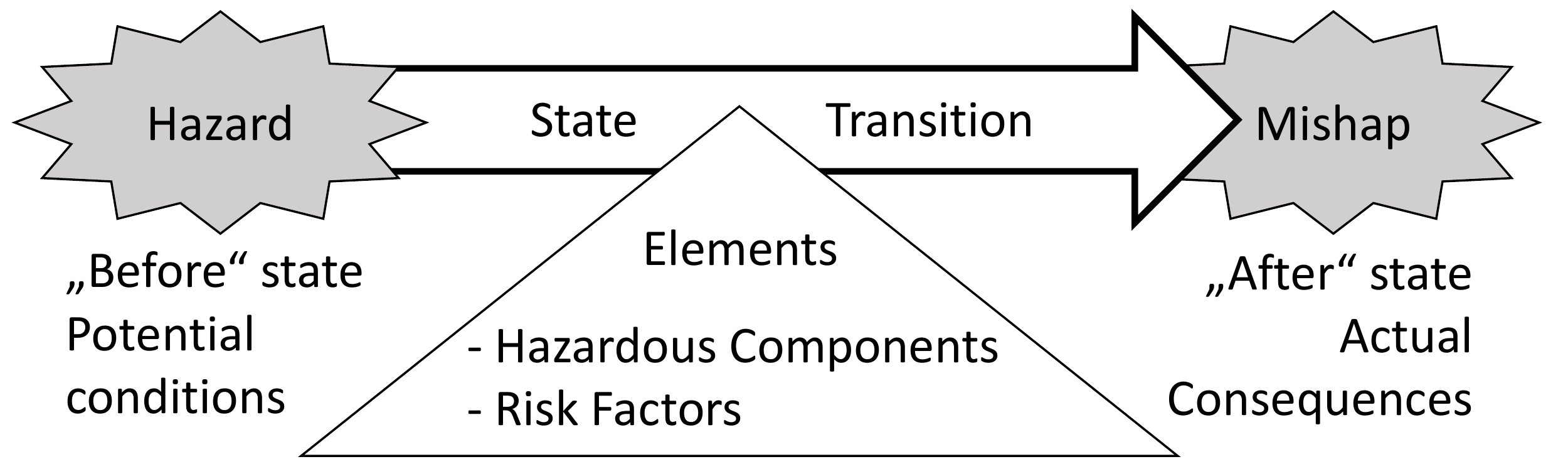}    % The printed column width is 8.4 cm.
	\caption{Coherence of hazard and mishap according to \cite{ericson_hazard_2005}} 
	\label{fig:hazard}
	\end{center}
\end{figure}

\begin{figure}[h]
	\begin{center}
	\includegraphics[width=0.4\textwidth]{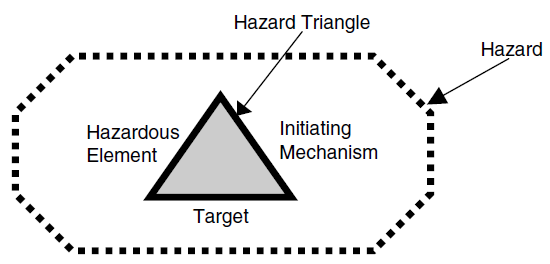}    % The printed column width is 8.4 cm.
	\caption{Hazard triangle \cite{ericson_hazard_2005}} 
	\label{fig:hazard_triangle}
	\end{center}
\end{figure}

The ISO~26262~standard defines a hazard as ``potential source of harm (1.56) caused by malfunctioning behavior (1.73) of the item (1.69)'', where \emph{harm} is ``physical injury or damage to the health of persons''.
\emph{Malfunctioning behavior} is defined as a ``failure (1.39) or unintended behavior of an item (1.69) with respect to its design intent'' \cite[Part 1]{ISO_26262_2001}.
According to Ericson \cite{ericson_hazard_2005} a hazard is a ``before'' state which transits to a mishap (or accident) through hazardous components and risk factors (Fig.~\ref{fig:hazard}).
Ericson further states that a hazard consists of \emph{hazardous elements}, \emph{initiating mechanisms} and \emph{targets} in the so called hazard triangle (Fig.~\ref{fig:hazard_triangle}).
%Following this definition and the anatomy of a hazard, the malfunctioning behavior is as a hazardous component, which causes a mishap from a hazardous event.
Following this definition and structure of a hazard, the malfunctioning behavior results from hazardous elements.
These elements cause mishaps because of initiating mechanisms (hazardous events).
The main objective of the HARA is to identify all components of the triangle and afterwards to remove or mitigate at least one of the components for each hazardous event in the functional safety concept.

Applying these definitions to vehicle guidance systems, the targets are other traffic participants and occupants of an automated vehicle.
There are three types of targets as ISO~26262~standard defines harm as damage or injury  to persons: other vehicles including occupants, cyclists and pedestrians.
The physical hazardous element is kinetic energy of the host vehicle and other traffic participants. 
This energy is influenced or regulated by functions, which are provided by the system.
Because there is no detailed technical solution available in the concept phase of development, functions are meant as an abstract description and not implemented functions.
The initiating mechanisms can be a malfunctioning behavior of the system or operation of the vehicle guidance system outside its functional system boundaries including the operational environment and weather.
Malfunctioning behavior itself cannot cause any harm without a scene including the environment of the vehicle which it is operating in.
To complete the representation of a hazard, the whole scene or at least the relevant components around the vehicle have to be described.
This information can be taken or generated from the item~definition and serve as input for the hazard analysis.

\section{Related Work}

The related work in the field of hazard analysis and hazard identification for automated vehicles is separated into two sections, because there is some recent work and many standards available.
The recent work provides an overview of current activities regarding the ISO~26262~standard and the traditional techniques section lists related standards.

\subsection{Recent work}

Different authors \cite{ross_funktionale_2013} \cite{ito_approach_2014} propose the scene-situation-matrix to identify hazardous events, but provide no method how to generate a complete set of relevant scenes for the item except expert knowledge in brainstorming.
This matrix can be used as a preliminary hazard identification method which contains a first guess of the system-risk.
Luo et al. \cite{selvaraj_modeling_2015} propose a modeling approach for safety case reasoning which covers hazardous events but do not mention how these events are generated.
Cuenot et al. \cite{cuenot_applying_2014} provide a model based-safety analysis for the system development phase in the ISO~262626~standard \cite[Part 4]{ISO_26262_2001}.
The system development phase is based on the concept phase and provides a detailed system design.
But it needs a functional safety concept to implement, which is derived from the HARA and the safety analysis in the concept phase.
Concluding the recent publications in the automotive domain, there seems to be potential towards systematic identification of hazards based on a functional system description.

\subsection{Traditional hazard analysis techniques}

\subsubsection{Hazard and Operability Analysis (HAZOP)} \hspace{\fill}

The Hazard and Operability Analysis (HAZOP) is a structured technique for examining a defined system, with the objective of identifying potential hazards in the system and identifying potential operability problems with the system \cite{IEC_61882_2001_HAZOP}.
Originally developed by the Institute of Chemical Industry (ICI) in the early 1970s \cite{lawley_operability_1974}, HAZOP studies were intended to analyze chemical plants.
Over the years HAZOP was applied to other fields like nuclear power plants, the petroleum industry, food and water industries and railways.
The key idea of HAZOP is to bring an interdisciplinary team together to assess proposed system deviations which are generated by combining an item with system parameters and specified guide words.
An example is a valve which generates too much flow, where \emph{valve} is the item, \emph{flow} is the parameter and \emph{too much} are the guide word(s).
The HAZOP method seems not to be applicable to vehicular systems without any modifications, because there are no guide words for the specific tasks of an automated vehicle available in the standard.
\newline

\subsubsection{Failure Mode and Effects Analysis (FMEA)} \hspace{\fill}

The Failure Mode and Effects Analysis (FMEA) is used to identify effects on the operability of the system caused by failures in subsystems, hardware components or system functions \cite{DIN_60812_2006_FMEA}.
Originally developed by the U.S. military in 1949 \cite{ericson_hazard_2005}, the standard was adapted for the automotive industry by the Ford Motor Company and was consolidated in 1993 by the Automotive Industry Action Group (AIAG).
Ericson \cite{ericson_hazard_2005} describes an analysis based on a functional model which describes \emph{what} the system does and not \emph{how} the functions are implemented in hard- and software.
This design FMEA (DFMEA) should be initiated right after the project start and during the concept phase.
DFMEA depends on design-responsibility, interfaces and interactions and the architecture of the system.
The architecture can be expressed with block diagrams, interface diagrams, functional diagrams, structure trees and schematic illustrations.
The item~definition provides the information to get a functional model for the item under investigation.
Based on this information the system functions (e.g. transforms, operates, contains) are expressed by requirements.
The analysis process identifies which components can fail and how this failure affects the requirements.
``The effects of the failure mode should be considered against the next level up assembly, the final product, and the end customer when known.''\cite{DIN_60812_2006_FMEA}
The difficulty for a driverless vehicle at the system level is to identify a complete or at least representative set of scenes where the effects can be investigated.
The (D)FMEA does not cover effects on the environment but can give hints how the system fails.
\newline

\subsubsection{Fault Tree Analysis (FTA)} \hspace{\fill}

The Fault Tree Analysis (FTA) is a deductive method for finding (basic) causes for unwanted events (malfunctions or malfunctioning behavior) in the system.
For this analysis a given effect of a system failure is analyzed in a Fault Tree to identify the components which can cause the effect \cite{IEC_61025_2006_FTA}.
The FTA can be used to assess which components in the system can cause the malfunctioning behavior in a hazardous event.
Fault Trees are derived from a detailed system design and are developed mainly for hardware analysis and a calculation of fault rates.
The functional safety concept, which is developed after the HARA, then shall mitigate or prevent failures in the identified system components.
\newline

\subsubsection{Event Tree Analysis (ETA)} \hspace{\fill}

The Event Tree Analysis (ETA) builds Event Trees, which are investigated according to the effect on the system (item) \cite{IEC_62502_2010_ETA}.
This method evaluates if the implemented safety mechanisms reduce or prevent the hazard from occurring.
The ISO~26262~standard demands from the hazard analysis and risk assessment that it should be processed without ``safety mechanisms intended to be implemented or that have already been implemented in predecessor items shall not be considered'' \cite[Part~3]{ISO_26262_2001}.
The ETA can be used to assess whether a functional safety concept is able to prevent or mitigate hazardous events but not to identify them.

\section{Approach for identification of potential hazardous events}

\begin{figure}[htb]
	\begin{center}
	\includegraphics[width=0.4\textwidth]{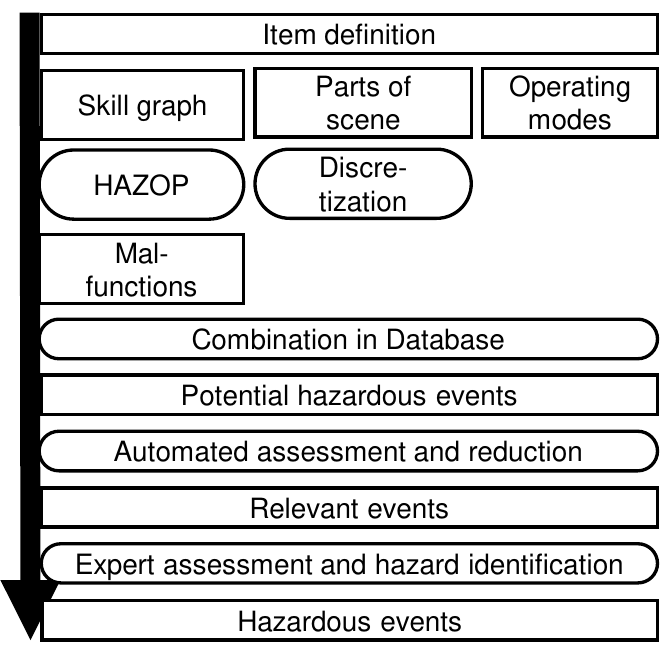}    % The printed column width is 8.4 cm.
	\caption{Proposed methodology for identification of potential hazardous events with work products (cornered) and process steps (rounded)} 
	\label{fig:process}
	\end{center}
\end{figure}

This section introduces the novel approach for identifying potential hazardous events (cf. Fig \ref{fig:process}).
The following parts show which entities were chosen to identify the events for the described unmanned protective vehicle.
A hazardous event consists of the current operating mode which is performed, a function with a specific malfunction and the current scene around the vehicle.
The next sections show how to identify these parts systematically and which values were chosen for the identification in the project aFAS.

\begin{figure*}[htpb]
	\begin{center}
	\includegraphics[width=17.9cm]{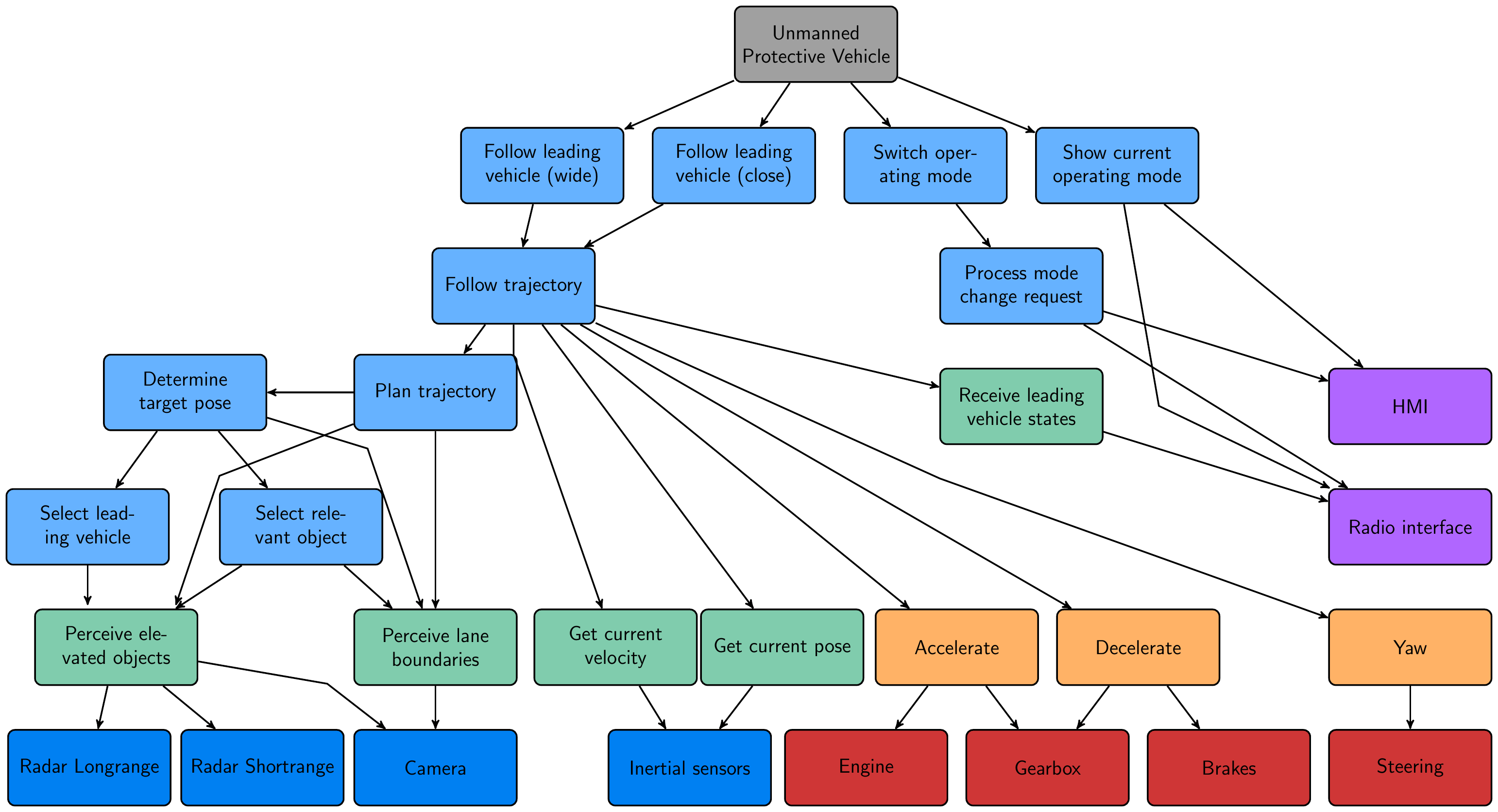}    % The printed column width is 8.4 cm.
	\caption{Skill graph of the unmanned protective vehicle categorized in main (grey), perception (green), planning (light blue), action (orange), sensors (blue), actuators (red) and input-output (purple), HMI: Human Machine Interface} 
	\label{fig:skillgraph}
	\end{center}
\end{figure*}

\subsection{Operating Modes}

The vehicle guidance system of the unmanned protective vehicle is planned to operate in four modes \cite{stolte_towards_2015}.
The first one is the \emph{Manual Mode}, in which the unmanned protective vehicle is controlled by a human and acts like a normal vehicle.
There is no operation of the vehicle guidance system in this mode.
When the unmanned protective vehicle is located at the working spot, the mode has to be switched to \emph{Safe Halt}.
In this mode the vehicle comes to a stop within the smallest possible distance or stands still.
This is used as the operational safe mode in case of a system failure or to start automated operation.
The automated driving functions are operated in \emph{Follow Mode} and \emph{Coupled Mode}.
Passing of acceleration and deceleration lanes is done in the \emph{Coupled Mode}, in which the unmanned protective vehicle drives with a very small gap between the two vehicles.
In this mode the leading vehicle's current states of actuators are sent to the unmanned protective vehicle via the radio interface to enable the operation in such close gaps.
This functionality is based on the work of the KONVOI \cite{deutschle_use_2010} project.
\emph{Follow Mode} is planned to be the main operation.
In this mode the unmanned protective vehicle has a gap of about 100~m to the leading vehicle.
It acts like an adaptive cruise control with stop and go feature and follows the hard shoulder like a lane keeping system.

\subsection{Functions and malfunctions}

According to Fig.~\ref{fig:conceptphase} the item~definition is the input for the HARA defining the unmanned protective vehicle in a functional way including system boundaries and operational environment.
From this informal document a skill graph, which describes the functionality with all dependencies, can be created.
In this case a skill is an abstract description of an activity which the system has to provide to fulfill the intended task \cite{Reschka2016}.
The item definition describes a system by the functional behavior and the system goals or mission.
Note that there should be no technical concepts in this part of the development process.
The system goals can then be divided into several sub-activities which have dependencies on each other. 
For example has the task \emph{following a lane} dependencies on \emph{perceiving a lane} and \emph{control the vehicles dynamic state}.
By modeling skills in a graph the system can be described from top-level goals or purpose over functional (and non-technical) dependencies to bottom sinks and sources which describe system boundaries.
Reschka et al. investigated the concept of skill graphs for appliance in vehicle guidance systems \cite{reschka_ability_2015}.
Note that the terms skills and abilities were interchanged.
Fig.~\ref{fig:skillgraph} shows the resulting skill graph, which was extracted from the item~definition.
For the purpose of describing vehicle guidance systems, the skills are separated into seven different categories:
\begin{itemize}
	\item System skill (grey) 
	\item Sensors (blue)
	\item Actuators (red)
	\item Input-Output (e.g. HMI) (purple)
	\item Perception skills (green)
	\item Planning skills (light blue)
	\item Action skills (orange)
\end{itemize}
The main or system skill describes the system itself and covers the overall functionality.
The underlying skills describe the system in a hierarchical way.
Beginning from the main tasks (following the lead vehicle and switching operating modes) in the different operating modes.
Skills are connected with arrows showing their dependencies to each other.
For example the skill \emph{select relevant object} has dependencies on a skill which perceives the object and another which perceives the hard shoulder boundaries to identify if an object is relevant or not.

To identify possible malfunctions we use the categorization of skills in combination with an adopted HAZOP methodology. % based on \cite{IEC_61882_2001_HAZOP}.
Hwang and Jo \cite{hwang_hazard_2013} used a modified \emph{HAZOP-R} (Railway) method in combination with a Preliminary Hazard Analysis (see \cite[chap. 5]{ericson_hazard_2005}) to identify hazardous events for a railway signaling system.
Trains mainly differ from vehicles because they are moving on rails and are coordinated by a central system.
In our adaption the HAZOP-\emph{item} is one skill of the categories perception, planning or action. 
The system parameters have to be defined according to the chosen skill.
For example \emph{Perceive objects} has parameters like \emph{relative position}, \emph{extent} and \emph{speed} for detected objects.
At this point we introduce keywords for the skill categories as follows:
%\vspace{-1mm}
\begin{itemize}
	\item Perception skills: No, nonexistent, erroneous, too large, too small
	\item Planning skills: Not relevant, Relevant \{\emph{parameter e.g. object}\} not, conflicting, physically not possible
	\item Action skills: Absent, wrong, unattended, too large, too small
\end{itemize}
We used the keywords to generate possible malfunctions for each skill in combination with a parameter of the skill.
For example, we used \emph{plan trajectory} (skill) planned a \emph{physically not possible} (keyword) \emph{turn rate} (parameter) in the trajectory.
The chosen categories fit for the reduced use case and system complexity of the unmanned protective vehicle.
Vehicle guidance systems with a wider functional range may need more detailed categories to declare usable guide words.

\subsection{Scenes}
\label{subsec:scenes}

\begin{figure}[h]
	\begin{center}
	\includegraphics[width=8cm]{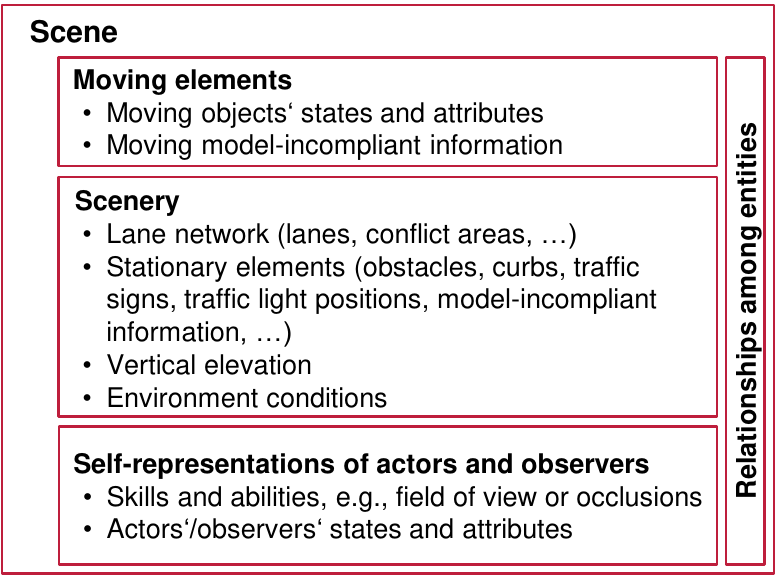}    % The printed column width is 8.4 cm.
	\caption{Parts of a scene according to \cite{ulbrich_scene_2015} }
	\label{fig:scene}
	\end{center}
\end{figure}

Ulbrich et. al \cite{ulbrich_scene_2015} recently reviewed many definitions of the term \emph{scene} and defined a consolidated definition as: \\ 
``A scene describes a snapshot of the environment including the scenery and dynamic elements, as well as all actors and observers self-representations, and the relationships among those entities. 
Only a scene representation in a simulated world can be all-encompassing (objective scene, ground truth).
In the real world it is incomplete, incorrect, uncertain, and from one or several observers points of view (subjective scene).'' \\
Thus, a scene consists of the three main parts \emph{dynamic~elements}, \emph{scenery} and \emph{self-representations of actors and observers} as shown in Fig.~\ref{fig:scene}.
For the purpose of defining scenes for the unmanned protective vehicle we chose the following entities:
\begin{itemize}
	\item Road infrastructure
	\item Road infrastructure width
	\item Road infrastructure curvature
	\item Traffic constellation
	\item Maximum velocity of moving traffic
	\item Weather conditions
	\item Object constellation on hard shoulder
	\item Driving state of the unmanned protective vehicle
	\item Malfunction as part of self-representation
\end{itemize}
The selected entities describe the operational environment for the reduced operating scenario of the unmanned protective vehicle and were identified by using information from the item definition.
Road infrastructures are described by a right and a left infrastructure in relation to the unmanned protective vehicle.
We added additional information to the scenes which exceed the functional system boundaries according to the item~definition.
This will be used to identify potential hazardous events, in which the vehicle has no malfunctions but is operated outside these boundaries.
These scenes can lead to mishaps without a malfunction or functional error and can be declared as a system boundary consideration.
The next challenge is to choose a level of discretization for each component of the scene.
As the hazard identification aims at generating top-level system hazards, we chose to use nearly binary discretization.
Choosing an appropriate level of discretization is a crucial step because too-detailed scenes can distort the risk assessment.
Very detailed scenes are more improbable than top level scenes and result in a lower exposure rating and this leading to a lower \emph{automotive safety integrity level} (ASIL) \cite[Part 3]{ISO_26262_2001} classification.
We choose a high level of discretization because the generated scenes shall provide a base for an expert team predicting the systems behavior in a given scene.
In our case the road infrastructures have \emph{solid markings}, \emph{dashed markings}, \emph{guardrails} and \emph{turf}.
The infrastructure width, infrastructure curvature and the weather conditions were set to \emph{valid} or \emph{invalid}.
The traffic constellation and the maximum velocity is chosen to \emph{moving traffic} with \emph{no limitation} according to the item~definition.
The object constellation on the hard shoulder can contain \emph{no object}, \emph{solid object} (like a car) or \emph{vulnerable object}.
The driving state of the unmanned protective vehicle is either \emph{stopped}, \emph{driving at 10~km/h} or \emph{driving at 80~km/h}.
This level of detail allows a very simplified and qualitative consideration of the operational scenes.

\subsection{Database}

After identifying all necessary components to describe events we have to generate the potential hazardous events.
For this step we created a SQL-database, where a permutation of all scenes is stored.
%The entity-relationship-model of the database is shown in Fig.~\ref{fig:erm}.
%\begin{figure*}[h]
%	\begin{center}
%	\includegraphics[width=16.8cm]{bilder/erm}    % The printed column width is 8.4 cm.
%	\caption{Entity-relationship-model of the potential hazardous events } 
%	\label{fig:erm}
%	\end{center}
%\end{figure*}
To analyze only relevant events, the database filtered scenes with following the constraints:
\begin{itemize}
	\item the function is not performed in the operating mode,
	\item multiple failures or functional system boundary exceeding or a combination of both exist,
	\item the malfunction is not relevant in scene (e.g. relevant object not considered in the scene where no objects are in place).
\end{itemize} 
In conformity with the ISO~26262~standard, multiple failures are not selected as relevant.
A functional system boundary exceedance is interpreted as a single failure for this contribution.
The relevant potential hazardous events then give an operating scene of the vehicle with a malfunction or system boundary exceeding in a defined operating mode as shown in Table \ref{tab:ex_event}.
\begin{table}[h]
	\caption{Example of a potential hazardous event}
	\renewcommand{\arraystretch}{1.4}
	\begin{tabular}{| p{0.215\textwidth}| p{0.215\textwidth}|}
	\hline 
	\textbf{Mode} & Follow Mode \\ 
	\hline
	\textbf{Function} & Select relevant object \\
	\hline
	\textbf{Malfunction} & Relevant object not considered \\ 
	\hline 
	\textbf{Road infrastructure} & Solid line (left) and turf (right) \\ 
	\hline 
	\textbf{Object constellation} & Vulnerable object \\
	\hline
	\textbf{Curvature, width and weather} & valid \\ 
	\hline 
	\textbf{Traffic constellation} & Moving traffic with no limitation \\ 
	\hline 
	\textbf{Driving state} & Driving at 10~km/h \\ 
	\hline 
	\end{tabular} 
	\label{tab:ex_event}
\end{table}

The scene in Table \ref{tab:ex_event} was classified as \emph{hazardous} because the vulnerable object, which can be either a human or an animal, would be injured if the vehicle does not stop. 
Based on these hazardous events, the top-level system hazards can be identified and afterwards be assessed in the risk~assessment step of the HARA.
 
\section{Results}

The number of scenes created for the unmanned protective vehicle according to the discretization from Section \ref{subsec:scenes} is 145 with 108 scenes classified as relevant.
This classification is based on whether there is no or only one system boundary exceedance.
We identified 16 functions with 37 malfunctions and the four operating modes for creation of the potential hazardous events.
Table \ref{tab:result} shows the numbers of generated and filtered events of the database.
\begin{table}[h]
	\caption{Numbers of generated and classified events}
	\renewcommand{\arraystretch}{1.25}
	\centering
	\begin{tabularx}{0.45\textwidth}{|*{4}{lXXX|}}
	\hline 
	\textbf{Mode} & \textbf{Events} & \textbf{Relevant} & \textbf{Hazardous} \\ 
	\hline 
	Manual Mode & 5328 & 373 & 238 \\ 
	\hline 
	Safe Halt & 5328 & 344 & 105 \\ 
	\hline 
	Follow Mode & 5328 & 377 & 170 \\ 
	\hline 
	Coupled Mode & 5328 & 368 & 237 \\ 
	\hline 
	\end{tabularx} 
	\label{tab:result}
\end{table}
The great decrease after filtering all generated events to relevant events can be explained by four factors shown in Table \ref{tab:factors}.
\begin{table}[htb!]\centering
	\caption{Major factors of filtering relevant events}
	\renewcommand{\arraystretch}{1.25}
	\begin{tabularx}{0.45\textwidth}{|*{4}{llll|}}
	\hline
	\multicolumn{4}{|p{0.45\textwidth-2\tabcolsep}|}{The function with a specific malfunction is not performed in operating mode} \\
	\hline
	Manual Mode & Safe Halt & Follow Mode & Coupled Mode \\
	2592 & 864 & 0 & 648 \\\hline
	\multicolumn{4}{|p{0.45\textwidth-2\tabcolsep}|}{Only one malfunction or system boundary exceedance is allowed} \\
	\hline
	Manual Mode & Safe Halt & Follow Mode & Coupled Mode \\
	864 & 720 & 1296 & 864 \\\hline
	\multicolumn{4}{|p{0.45\textwidth-2\tabcolsep}|}{The combination of malfunction and scene is (physically) invalid} \\
	\hline
	Manual Mode & Safe Halt & Follow Mode & Coupled Mode \\
	1331 & 666 & 666 & 666 \\\hline
	\multicolumn{4}{|p{0.45\textwidth-2\tabcolsep}|}{Scene is not relevant for operating mode} \\
	\hline
	Manual Mode & Safe Halt & Follow Mode & Coupled Mode \\
	0 & 2664 & 2664 & 2664 \\\hline
	\end{tabularx}
	\label{tab:factors}
\end{table}
The first point of automated reduction is decided by whether a specific function is operated in a certain operating mode.
In Manual Mode a roadworker is driving the unmanned protective vehicle like a normal truck, thus only acting skills are used in Manual Mode and the number of relevant events is significantly reduced.
Safe~Halt and Coupled~Mode are using a reduced set of system functionality as both do not perceive lane boundaries and Safe Halt not even objects. \\
Second influence on the number of relevant events is the selection of only one malfunction or system boundary exceedance.
This criterion comes from the ISO~26262~standard where only single point of failures are considered.
The number of relevant events will increase significantly, if two or more malfunctions are evaluated. \\
Due to automated generation of the events without any physical or formal modeling, there are plenty of events which are (physically) not possible or just not meaningful.
For example is an event with the malfunction \emph{existing object not recognized} in a scene where \emph{no object} is located on the hard shoulder not meaningful. \\
Last of all some scenes are not possible for some operating modes. 
The reduction of 2664 events in all automated operating modes is explained by scenes where the velocity of the unmanned protective vehicle is at 80~km/h.
This state is only relevant for Manual Mode as the maximum velocity in the automated modes is limited to 10~km/h. \\
The relevant events then were assessed by a team of experts as to whether they are hazardous or not.
We can not provide proof of completeness since the development process is still on-going and to the best of our knowledge there is no measure for completeness.
For validation purposes, a next step is to compare the resulting hazards from this approach with the identified hazards by a team of experts.

\section{Conclusion and Future work}

The generated potential hazardous events and the (plausibility) filtering in a database result in a systematic way to identify top level system hazards for an unmanned protective vehicle with a very limited use case but the first planned unmanned operation in public traffic.
A drawback of the proposed method is that there are several events leading to the same hazards and the same ASIL classification.
For example the infrastructure width does not have any impact on how the actuating skills, like accelerating the vehicle, perform.
This topic can be addressed by using equivalence classes for parts of a scene towards selected skills and with this reduce the total number of events by not loosing information about critical events.
A difficulty in choosing equivalence classes is to prove that no potential hazardous event is omitted.
These equivalence classes could be identified by having a look at skill categories with regard to single parts or categories of the overall scene definition.
Another focus is to extend this method for vehicle guidance systems with a wider use case, like the project \emph{Stadtpilot} \cite{nothdurft_stadtpilot:_2011}.
The first step for this purpose is to generate an item~definition and to describe the operational environment in inner cities.
Due to the situational complexity of this environment the functions for evaluating and reducing the total number of events must be extended because there is a huge amount of possible events.

% use section* for acknowledgment
\section*{Acknowledgment}

We would like to thank our partners from the project consortium consisting of MAN Truck - Bus AG (consortium leader), TRW Automotive GmbH (ZF TRW), WABCO Development GmbH, Robert Bosch Automotive Steering GmbH, Hochschule Karlsruhe, Technische Universit\"at Braunschweig, Hessen Mobil - Road and Traffic Management, and BASt - Federal Highway Research Institute for their support. 
The project is partially funded by the German Federal Ministry for Economic Affairs and Energy.

% trigger a \newpage just before the given reference
% number - used to balance the columns on the last page
% adjust value as needed - may need to be readjusted if
% the document is modified later
%\IEEEtriggeratref{8}
% The "triggered" command can be changed if desired:
%\IEEEtriggercmd{\enlargethispage{-5in}}

% references section

% can use a bibliography generated by BibTeX as a .bbl file
% BibTeX documentation can be easily obtained at:
% http://mirror.ctan.org/biblio/bibtex/contrib/doc/
% The IEEEtran BibTeX style support page is at:
% http://www.michaelshell.org/tex/ieeetran/bibtex/
\bibliographystyle{IEEEtran}
% argument is your BibTeX string definitions and bibliography database(s)
\bibliography{biblio}

%
% <OR> manually copy in the resultant .bbl file
% set second argument of \begin to the number of references
% (used to reserve space for the reference number labels box)
%\begin{thebibliography}{1}

%\bibitem{IEEEhowto:kopka}
%H.~Kopka and P.~W. Daly, \emph{A Guide to \LaTeX}, 3rd~ed.\hskip 1em plus
%  0.5em minus 0.4em\relax Harlow, England: Addison-Wesley, 1999.

%\end{thebibliography}

% that's all folks
\end{document}